\documentclass[onecolumn,prd,superscriptaddress,nofootinbib,notitlepage]{revtex4-1}
\usepackage{latexsym}
\usepackage{amsmath,amsfonts,graphicx,accents}
\usepackage{amsbsy}
\usepackage{hyperref}
\usepackage{mathrsfs}
\usepackage{color}
\usepackage{psfrag}
\usepackage{enumerate}
\usepackage{amsmath,amssymb,calc,amsfonts}
\usepackage{latexsym}
\usepackage[utf8]{inputenc}
\usepackage[percent]{overpic}
\usepackage[autostyle=false, style=english]{csquotes}
\MakeOuterQuote{"}
\DeclareFontFamily{U}{rsfs}{}         
\DeclareFontShape{U}{rsfs}{m}{n}{<5> rsfs5 <6><7> rsfs7          %
  <8><9><10><10.95><12><14.4><17.28><20.74><24.88> rsfs10}{}     %
\DeclareMathAlphabet{\mathfs}{U}{rsfs}{m}{n}                     %
\definecolor{indiagreen}{rgb}{0.07, 0.53, 0.03}
\def\beq{\begin{eqnarray}}
\def\eeq{\end{eqnarray}}

\def\={\stackrel{\Delta}{=}}

\begin{document}
\title{Physical Process First Law and the Entropy Change of Rindler Horizons}

\author{T. K. Safir}\email{stkphy@gmail.com}
\affiliation{Department of Physics, T.K.M College of Arts and Science, Kollam, Kerala 691005, India.} 
\affiliation{Department of Physics, National Institute of Technology Karnataka, Surathkal  575025, India}
\author{C. Fairoos}\email{fairoos.phy@tkmcas.ac.in}
\affiliation{Department of Physics, T.K.M College of Arts and Science, Kollam, Kerala 691005, India.} 

\author{Deepak Vaid}
\email{dvaid79@gmail.com}
\affiliation{Department of Physics, National Institute of Technology Karnataka, Surathkal  575025, India}

\begin{abstract}
The physical process version of the first law can be obtained for bifurcate Killing horizons with certain assumptions. Especially, one has to restrict to the situations where the horizon evolution is quasi-stationary, under perturbations. We revisit the analysis of this assumption considering the horizon perturbations of Rindler horizon by a spherically symmetric object. We demonstrate that even if the quasi-stationary assumption holds, the change in entropy, in four space-time dimensions, diverges when considered between asymptotic cross-sections. However, these divergences do not appear in higher dimensions. We also analyze these features in the presence of a positive cosmological constant.  In the process, we prescribe a recipe to establish the physical process first law in such ill-behaved scenarios.
\end{abstract}

 \maketitle
\section{Introduction}
Black holes are perhaps the simplest non-trivial solutions of Einstein's field equations. However, the implications of its geometry are both interesting and intriguing. The presence of a horizon prevents classical information, inside the horizon, from reaching an asymptotic observer. Due to this fact, the black hole horizon can be treated as an inner boundary of spacetime. It is this feature that gives rise to a relation between the infinitesimal changes in mass, charge, angular momentum of a black hole and the change of its horizon area, akin to the first law of thermodynamics \cite{Bardeen:1973gs}. These infinitesimal changes are on the space of stationary black hole solutions, and this relation is referred to as the \textit{stationary state} version of the first law.
Further, it was argued that black holes, in general relativity, can be endowed with an entropy which is proportional to the area of the horizon \cite{Bekenstein:1972tm,Bekenstein:1973ur}. The result of \cite{Hawking:1971vc,Hawking:1974sw} that black holes radiate, quantum mechanically, like
a black body, at a temperature equal to that of its surface gravity, further lends support to the fact that black holes indeed behave like thermodynamic objects. These further raise important questions regarding the quantum origin of this entropy and motivates the quest for a quantum theory of gravity. However, the first law, which is purely of classical origin, remains an important aspect to explore.\\

Black holes, as found in nature, are far from being appropriately described by global stationary black hole solutions. They evolve due to infalling matter, and the area of the horizon changes with time. Unlike the infinitesimal change in the stationary state version of the first law, which is a change in the space of solutions, this is an evolution in time due to the process of matter falling into the black hole. Consequently, a different version of the first law holds and appropriately describes this situation. In this version one relates the time evolution of the entropy to the matter influx across the horizon \cite{Hawking:1972hy,CarterReview,wald1994quantum} and is therefore called as the "\textit{physical process}" version of the first law (PPFL). As will be explicit, PPFL can be locally characterized, and therefore, it also holds for a  wider class of horizons, which does not require the specification of the asymptotic structure. As a result, PPFL holds in the context of Rindler \cite{Jacobson:2003wv}, or for that matter, any bifurcate Killing horizon \cite{Amsel:2007mh}. \\  

 If a stationary horizon is perturbed by some matter stress-energy tensor $T_{\mu\nu}$ and, once the black hole settles down to another stationary state, then PPFL provides a mathematical expression for the change in horizon area $A_H$ as,
\begin{equation}
\frac{ \kappa }{ 8 \pi } \Delta A_H = \int_{\mathcal{H}} T_{\alpha \beta} \chi^\alpha \chi^\beta dA~dt,
 \end{equation}
 where $ dA$ is the area element of the horizon cross-section, $t$ is the Killing parameter associated with the horizon-generating Killing vector $\chi^\mu$ and $\kappa$ denotes the surface gravity of the unperturbed horizon. Much like the equilibrium version of the usual first law of thermodynamics, the above relation will fail to hold, if considered between two nonequilibrium states, due to dissipative effects. This would mean that the process of evolution of an initial equilibrium state has taken place in a non-quasi-static manner. Then one can ask, what are the processes that are quasi-static such that PPFL continues to hold. It is important to note that while deriving the above relation, by a straightforward integration of the Raychaudhuri equation, there are two assumptions made. First, the process of horizon evolution must be quasi-static so that the terms which are of higher order in expansion and shear of the horizon can be neglected. Second, upon perturbation and in the course of evolution, no additional generators are added to the horizon. When an object falls into the black hole causing the evolution of the horizon, the interesting question is, what are the restrictions on the parameters of the object such that the above assumptions remain valid. This query was first examined by \cite{Thorne,Suen:1988kq} and further extended for general perturbations by \cite{Amsel:2007mh} and \cite{Bhattacharjee:2014eea}.\\
 
A necessary condition such that the above assumptions remain valid during the evolution, is the avoidance of caustic formation along the horizon, a situation where the expansion becomes negative infinity. Not only does the formation of caustic violates the first assumption, but it also indicates the addition of new generators to the horizon. In $D=4$, if a black hole of mass $M$ is perturbed by a spherically symmetric object of mass $m$ and radius $r$ then in order to avoid caustic formation, along the horizon, the radius of the object ($r$) must satisfy $ r > 2 \sqrt{2 M m} $ \cite{Thorne}. Such a constraint on the size of the perturbing matter has also been obtained for Rindler horizon in arbitrary dimensions \cite{Amsel:2007mh} as well as for perturbing matter which is charged or rotating\cite{Bhattacharjee:2014eea}.\\

In this work, we demonstrate that even if such a constraint holds, the change in horizon entropy diverges, when considered between asymptotic cross-sections as pointed out in \cite{Jacobson:2003wv}. However, we present an interesting observation that such divergences are absent for dimensions greater than 4. The question now is whether one can modify the PPFL to account for such situations or in more general situations where caustics `do' form to the future of the bifurcation surface. The hint comes from a modification of PPFL suggested in \cite{Chakraborty:2017kob}, where it was shown that one could choose arbitrary cross-sections as the initial and final slices, at the cost of introduction of an extra term in the first law expression. Hence we show that even if a caustic forms to the future of the bifurcation surface, one can take our initial slice beyond that point and write a modified PPFL. Finally, we discuss the same problem in the presence of a positive cosmological constant.\\

\section{Evolution of the horizon and the formation of caustic}\label{formalism}
In this section, we briefly review the derivation of PPFL as well as the conditions on the perturbation strength,  along the lines of \cite{Amsel:2007mh} and \cite{Bhattacharjee:2014eea}, such that caustic is avoided to the future of the bifurcation surface. Consider a space-time with a bifurcate Killing horizon. Suppose some energy flux falls across an initially stationary bifurcate Killing horizon. Due to the teleological nature of the event horizon, it starts expanding even when there is no flux across it and finally settles down to another stationary state once the passage of all matter has ceased. The corresponding change in the expansion $\theta$, of the horizon, is governed by the Raychaudhuri equation for null-congruences, associated with the horizon generating Killing vector $ \chi^\mu$,
\begin{equation}\label{Ray}
\frac{ d \theta }{ dt } = \kappa \theta  - \frac{1}{2} \theta^2 - \sigma_{\mu \nu } \sigma^{\mu \nu}- R_{\alpha \beta} \chi^\alpha \chi^\beta.
\end{equation}
We have parametrized the geodesics by the Killing time $t$ and $t= -\infty$ corresponds to the bifurcation surface. Further, $\kappa$ is defined as $ \chi^\mu \nabla_\mu \chi^\nu = \kappa \chi^\nu$ and $\sigma_{\mu\nu}$ represents the shear of this congruence. The basic assumption one takes while deriving the physical process version of the first law is that the process of horizon evolution is quasi-stationary (quasi-static in the thermodynamic sense), such that one can neglect terms which are higher-order in $\theta$ and $\sigma_{\mu\nu}$, in Raychaudhuri equation. Under this assumption, Eq. (\ref{Ray}) becomes,
\begin{equation}
-\frac{d \theta}{dt} +\kappa \theta = \mathcal{S}(t),
\end{equation}
where  $\mathcal{S}(t) = 8\pi T_{ \alpha \beta} \chi^\alpha \chi^\beta $ is the perturbing energy flux. One exploits the Green's function technique to solve this differential equation. Further, using the expression $\theta=\frac{1}{ \Delta A } \left( \frac{ d(\Delta A) }{ dt } \right) $, one gets the final result as,
\begin{equation}
\frac{ d(\Delta A) }{ dA } = \frac{ 8 \pi }{ \kappa} \int_{-\infty}^{\infty} dt \ T_{ \alpha \beta } \chi^{\alpha} \chi^{\beta},
\end{equation}
where $A$ denotes the area of horizon cross-section. The right-hand side of the above equation can be identified as the Killing energy ($E_\chi$) associated with $\chi^\mu$, which is the amount of energy crossing the horizon. Thus we have the first law,
\begin{equation}
 \frac{\kappa \Delta A}{8\pi} = \Delta E_\chi,
\end{equation}
where $\Delta A$ represents the difference between the area of the perturbed horizon in the asymptotic future and the initial area of the horizon before perturbation. The lower limit of the integration was taken to be the bifurcation surface ($t=-\infty$). As we have discussed before, the crucial assumption one has to make while deriving the first law is the quasi-stationarity of the horizon evolution. This statement about quasi-stationarity can be quantified by calculating a threshold value of $\theta$, beyond which caustics set in. To arrive at an expression for the threshold value, consider the homogeneous version ($\mathcal{S} = \sigma^{2} = 0 $) of the Raychaudhuri equation,
\begin{equation}
\left(-\frac{d}{dt} + \kappa \right) \frac{\theta}{2 \kappa} - \kappa\left(\frac{\theta}{2 \kappa}\right)^2 = 0.
\end{equation}
Solving for $ \left({\theta}/{(2 \kappa)}\right)$ we obtain,
\begin{equation}
\frac{\theta}{2 \kappa} = \frac{1}{\left(1+\left(\frac{2 \kappa}{\theta_0}-1\right)e^{\kappa(t_0-t)}\right)},
\end{equation}
where $\theta_0$ is the expansion of the horizon at some time $t_0$ before the driving force acts. We can clearly see that, if ${\theta_0}/{(2 \kappa)} > 1$, then $\theta$ increases to the past and becomes infinite at some finite time $t < t_0$ \cite{{Thorne}}. This means, if the expansion becomes greater than $2 \kappa$ at any instance during the evolution, then the Raychaudhuri equation implies that the horizon would develop a caustic at a finite earlier time. Evidently, our earlier approximation in deriving the first law breaks down in this case. Our goal is to see what this condition means in terms of the parameters of the perturbing object. Before proceeding, we outline a few equations related to the horizon evolution.\\

The complete description of the horizon evolution is governed by two equations. The Raychaudhuri equation (Eq. (\ref{Ray})), which says how the expansion changes along the horizon Killing parameter and, the tidal force equation which gives the evolution of the shear $\sigma_{\mu \nu}$, i.e., 
\begin{eqnarray}\label{tidal}
\frac{d \sigma_{\mu \nu}}{dt} = (\kappa-\theta)\sigma_{\mu \nu} - \sigma_{\mu \sigma}\sigma^\sigma~_{\nu} + \frac{1}{2}\sigma^2 h_{\mu \nu} + \left(2 \sigma_{\mu \sigma} + \theta h_{\mu \sigma}\right)\sigma^\sigma~_{\nu} -\varepsilon_{\mu \nu},
\end{eqnarray}
where $\varepsilon_{\mu \nu} = h^\alpha_\mu h^\beta_\nu C_{\alpha\lambda\beta\sigma}\chi^\lambda \chi^\sigma$ is the electric part of the Weyl tensor $C_{\alpha\lambda\beta\sigma}$, and, $h_{\mu \nu}$ is the projection operator onto the space-like cross-sections of the horizon and is defined as,
\begin{equation}
h_{\mu \nu} = g_{\mu\nu} + \chi_\mu l_\nu + \chi_\nu l_\mu.
\end{equation}
Here $l_\mu$ is an auxiliary null vector, defined as  $l_\mu \chi^\mu =-1$. If the horizon perturbation is weak, then both Eq. (\ref{Ray}) and Eq. (\ref{tidal}) will get simplified significantly. We take the tidal field, which sources the evolution of shear, to be of first-order in a small dimensionless parameter $m\kappa$. Here, $m$ denotes the mass of the perturbing object, and $\kappa$ is the surface gravity of the unperturbed horizon. Truncating both equations to lowest order in the perturbation parameter, we get,
\begin{equation}\label{ti}
-\frac{d \sigma_{\mu \nu}}{dt} +\kappa \sigma_{\mu \nu} = \varepsilon_{\mu \nu},
\end{equation}
and,
\begin{equation}\label{Ra}
-\frac{d \theta}{dt} + \kappa \theta = \mathcal{S}(t) + \sigma^2.
\end{equation}
The tidal field sources the evolution of the shear $\sigma_{\mu\nu}$ according to Eq. (\ref{ti}) and this $\sigma_{\mu\nu}$ along with the non-gravitational energy flux determines the horizon expansion through Eq. (\ref{Ra}). The above equations can be solved to get,
\begin{equation}\label{sigma}
\sigma_{\mu\nu}(t) = \int_{-\infty}^{\infty} \varepsilon_{\mu \nu}(t')e^{\kappa(t-t')}\Theta(t'-t)dt',
\end{equation}
\begin{equation}\label{theta}
\theta(t) = \int_{-\infty}^{\infty} \left(\mathcal{S}(t') + \sigma^2(t')\right)e^{\kappa(t-t')}\Theta(t'-t)dt'.
\end{equation}
Having laid out the necessary tools, we now proceed to calculate the conditions on the parameters of the perturbing object so that the PPFL remains valid. The investigation was first carried out in \cite{Thorne,Suen:1988kq}, where authors considered the perturbation of a Kerr black hole by a freely falling object. The calculations were done by a justifiable approximation of a Kerr horizon by a Rindler horizon (RH), the horizon perceived by an accelerating observer in flat spacetime. They arrived at the following condition on the radius ($r$) of the perturbing object so that the PPFL remains valid.
\begin{equation}\label{threshold theta}
r > 2\sqrt{\frac{m}{2\kappa}},
\end{equation}
where $\kappa$ is the surface gravity of the black hole and $m$ is the mass of the perturbing object. If the radius of the perturbing object, falling into the horizon, is less than that of the threshold value give by Eq. (\ref{threshold theta}), then caustic will form along the horizon invalidating the approximations made while deriving PPFL. This result, however, cannot be adopted for the case of Rindler horizon by taking $\kappa \rightarrow 0$ limit. This was pointed out later in \cite{Amsel:2007mh} and a result that holds for general bifurcate Killing horizons, in spacetime dimension $D\geq3$, was obtained. Further, in \cite{Bhattacharjee:2014eea}, the authors considered situations where a freely falling charged or rotating object perturbs a Rindler horizon in $D=4$.

\section{Perturbation of Rindler Horizon by a freely falling object}
In this section, we briefly recap the problem regarding the validity of PPFL by considering a horizon perturbation by a freely falling object. We obtain a constraint on the size of the perturbing object so that caustic formation is avoided along the horizon. Further, we explicitly show how to establish the first law even if caustic forms along the horizon at some point in time. To calculate horizon expansion of a black hole, we approximate the black hole horizon by a horizon perceived by an accelerating observer in flat space-time (Rindler horizon). This is very reasonable since we restrict the region of study very close to the event horizon, which can be approximated to be Rindler. Then, we can obtain the corresponding results for the perturbation of the black hole horizon, as explained in Appendix (\ref{one}). From now on, we will be considering the evolution of the Rindler horizon only.

\subsection{Caustic avoidance of Rindler horizon in Four spacetime dimension}\label{boud_cal}
\begin{figure}[h]
\center
\includegraphics[width=12cm]{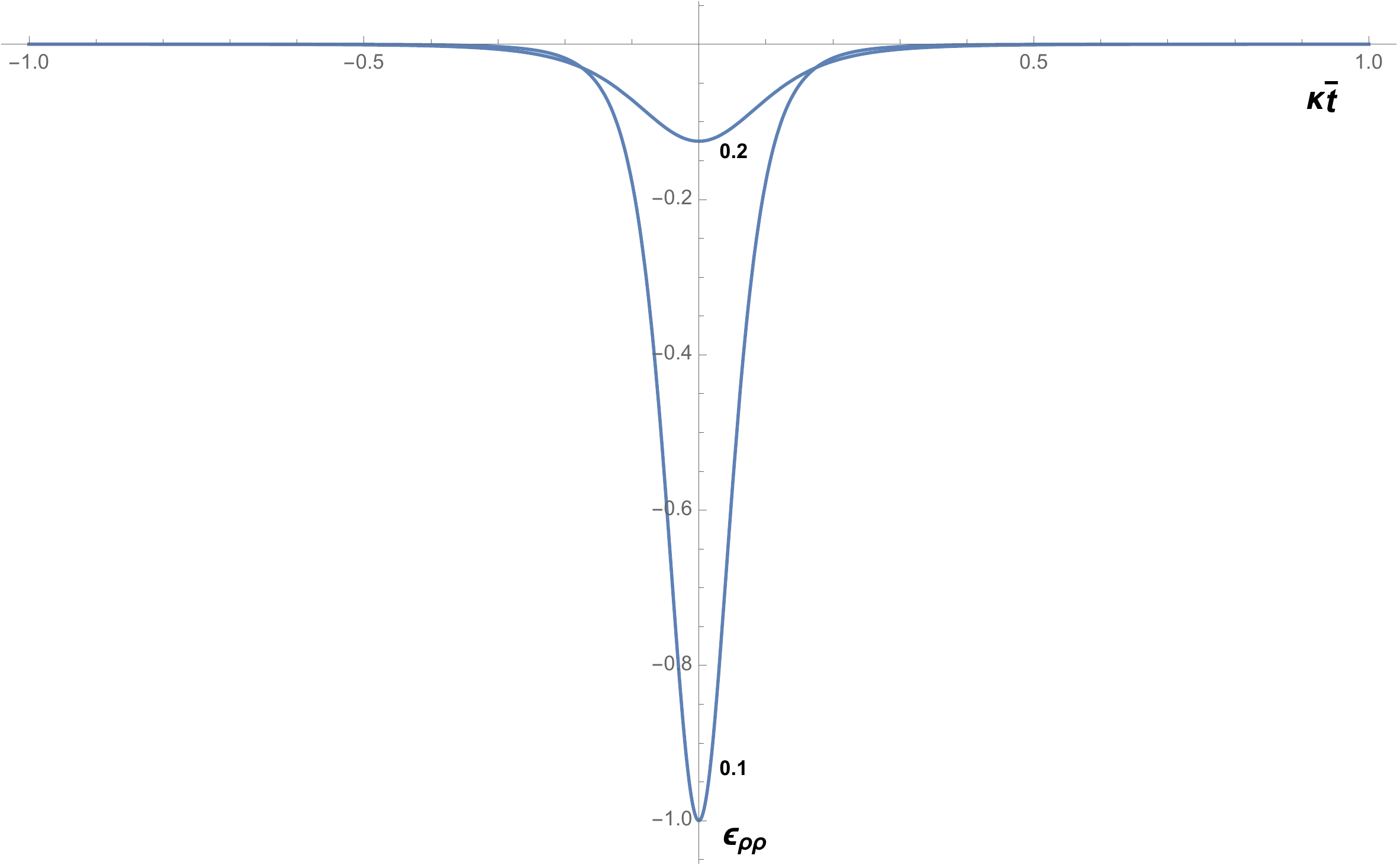}
\caption{\small{Behaviour of the Weyl tensor against $\kappa \bar{t}$ for two different values of $\frac{\rho}{z_0}$. One can see that the profile tends to look like a delta function as $\frac{\rho}{z_0} \to 0$.}}
\label{fig.delta}
\end{figure}

Consider a spherically symmetric object falling across a Rindler horizon. We assume that the mass of the perturbing matter is very small so that one can neglect the back-reaction effects. Also, we neglect terms which are higher-order in $m$ throughout the calculations, where $m$ is the mass of the perturbing object. In Minkowski coordinates ($T,Z,x,y$), the trajectory of the freely falling object is given as, $x=y=0,$ and $ Z=z_0$. The perturbing object is characterized by the solution of the linearized Einstein's field equations. In isotropic coordinates, the perturbing metric is given as,
\begin{eqnarray}
ds^2 &=& -\left(1-\frac{ 2 m }{ \sqrt{ \rho^2 + (Z-z_0)^2 } }  \right) dT^2 
\\ \nonumber &+& \left( 1 + \frac{ 2 m }{ \sqrt{ \rho^2 + (Z-z_0)^2 } }\right)\left( dZ^2 + d \rho^2 + \rho^2 d \theta^2 \right),
\end{eqnarray}
where $\rho^2 = x^2+y^2$. We calculate the non-zero components of the electric part of the Weyl tensor. On the horizon ($T=Z$) we have,
\begin{eqnarray}\label{exactweyl}
\varepsilon_{\rho \rho} = - \frac{1}{\rho^2} \varepsilon_{\theta \theta} 
=\left(\frac{-3 m \rho^2 \kappa^2 Z^2}{\left(\rho^2+(Z-z_0)^2\right)^{\frac{5}{2}}}\right) + O(m^2).
\end{eqnarray}
Further, we express the above relation in terms of advanced Killing time which is related to the Minkowski time co-ordinate as,
\begin{equation}
t = \frac{1}{\kappa} \ln\left(\frac{\kappa}{2}(T+Z)\right).
\end{equation}
 Along the horizon we have,
\begin{equation}
T =Z= z_0 e^{\kappa \bar{t}},
\end{equation}
where $\bar{t} = t-t_0$ is the shifted Killing time and $t_0$ denotes the time at which matter crosses the horizon. For $\kappa \bar{t} \ll 1$ one can expand the above exponential. Thus the electric part of the Weyl tensor becomes,
\begin{eqnarray}\label{complete_exprn_1}
\varepsilon_{\rho \rho} &=& -\frac{3 \rho^2 \kappa^2 z_0^2 m }{\left(\rho^2+z_0^2\kappa^2\bar{t}^2\right)^{\frac{5}{2}}} .
\end{eqnarray}
One can see that the maximum contribution to the electric part of the Weyl tensor comes from $\bar{t} =0$. We assume ${\rho}/{z_0} \ll 1$. Now, the time dependence of the above function can be effectively described by a delta function peaked at $\bar{t} =0$ \cite{{Thorne}}. This can be understood by plotting the behavior of the Weyl tensor as a function of $\kappa \bar{t}$. One can easily see from the figure (\ref{fig.delta}) that the Weyl tensor behaves like a delta function in the region ${\rho}/{z_0} \ll 1$. Therefore, we express the electric part of the Weyl tensor as,
\begin{equation}\label{delta_apprxmn_1}
\varepsilon_{\rho \rho}(\bar{t}) = -\frac{4 m \kappa z_0}{\rho^2} \delta(\bar{t}).
\end{equation}
\\
The approximation of Eq. (\ref{complete_exprn_1}) by Eq. (\ref{delta_apprxmn_1}) is justified as follows. It is clear that the maximum expansion obtained using Eq. (\ref{delta_apprxmn_1}) will always be greater than the one obtained from Eq. (\ref{complete_exprn_1}). Hence, if the maximum value of expansion obtained using Eq. (\ref{delta_apprxmn_1}) satisfies the condition of caustic avoidance, it will hold true for the expansion resulting from Eq. (\ref{complete_exprn_1}). Now, we calculate the horizon shear using Eq. (\ref{sigma}). We get,
\begin{equation}\label{sigma_0}
\sigma_{\rho \rho} = -\frac{1}{\rho^2} \sigma_{\theta \theta}= -\frac{4 m \kappa z_0}{\rho^2} e^{\kappa \bar{t}} \Theta(-\bar{t}).
\end{equation}
This expression says that the horizon shear vanishes once the perturbing object crosses the horizon ($\bar{t} =0$). We now calculate the expansion of the horizon using Eq. (\ref{theta}). The integration can be done easily, resulting in the following expression.
\begin{eqnarray}\label{theta_BH_1}
 \theta(\bar{t}) &=& \frac{2}{\kappa}  \left(\frac{4 m \kappa z_0}{\rho^2} \right)^2 \left(1-e^{\kappa \bar{t}}\right)e^{\kappa \bar{t}}\Theta(-\bar{t}).
\end{eqnarray}
 Similar to the shear, the expansion of the horizon also vanishes after the object has crossed the horizon. It is to be noted that, the expression above determines the expansion of those geodesics which pass outside the perturbing object. Our goal here is to find the allowed sizes of the perturbing object so that no caustic forms along the horizon, ensuring the validity of the physical process first law. Note that the expression, i.e., Eq. (\ref{theta_BH_1}), describes the expansion of a Rindler horizon when a spherically symmetric charged matter falls freely onto it. One can also find the corresponding expression for a black hole horizon by replacing $z_0$ with $1/\kappa$, where $\kappa$ is the surface gravity of the black hole (Appendix (\ref{one})). If we approximate $\rho$ by the radius $r$ of the perturbing object \footnote{This approximation is reasonable since we have shown that the electric part of the Weyl tensor behaves as a delta function peaked at $\bar{t} = 0$, which is the time at which perturbing matter crosses the horizon. At this point the radial co-ordinate $\rho$ of the matter is essentially its radius $r$.}, the condition for caustic avoidance along the horizon ($\theta_{\text{max}}<2\kappa$) reads,
\begin{equation}\label{first_condition}
r > 2\sqrt{\frac{m}{2 \kappa}}.
\end{equation}
This result has been reported in \cite{Thorne,Amsel:2007mh,Suen:1988kq}.

\subsection{Caustic avoidance for Higher dimensional Rindler Horizon}
In dimensions higher than 4, one can calculate the bound on the radius of the perturbing object by following the same steps and approximations as in section (\ref{boud_cal}). The metric for the perturbing object in $ (n+2)$ dimensional space-time is given by,
\begin{eqnarray}
ds^2 &=& -\left(1-\frac{C m }{ \left(\sqrt{\rho^2+(Z-{z_0})^2}\right)^{(n-1)}}\right) dT^2 \\ \nonumber &+& \left( 1+ \frac{C m}{ (n-1)\left(\sqrt{\rho^2+(Z-z_0)^2}\right)^{(n-1)}}\right)\left(dZ^2+ d\rho^2 + \rho^2 d\Omega_{n-1}^2\right),
\end{eqnarray}

where $ \rho^2 = \sum_{i=1}^n x_i^2$ and $x_i$'s are the Minkowski co-ordinates. The constant $C = \frac{16 \pi}{n \Omega_n}$, with $ \Omega_n$ being the volume of $S^n$. The non-zero components of the electric part of the Weyl tensor are calculated along the horizon ($Z-T=0$) as, 
\begin{gather}\label{Weyl_n}
\varepsilon_{\rho \rho} =  - \frac{(n-1)}{g_{\theta_i \theta_i}}\varepsilon_{\theta_i \theta_i} = -\frac{ (n-1)(n+1)C m \kappa^2 Z^2 \rho^2}{2\left(\sqrt{\rho^2+(Z-z_0)^2}\right)^{(n+3)}}.
\end{gather}
Proceeding with the same analysis as before, the formation of caustic is avoided if the radius of the object satisfies the following relation.
\begin{equation}\label{Hrbound}
r > \left( \frac{4 \pi \sqrt{n-1}  m z_0}{\Omega_{n-1}}\right)^{\frac{1}{n}}.
\end{equation}

This result was obtained in \cite{Amsel:2007mh}. We would like to extend the analysis further. Suppose the radius of the perturbing object does not satisfy Eq. (\ref{first_condition}), then caustic may develop along the horizon at some finite earlier time. In such situations, the PPFL becomes invalid. However, we explicitly show that one can always recover the first law for a certain interval of horizon evolution according to \cite{Chakraborty:2017kob}, in which the authors have obtained a first law expression for arbitrary horizon slices. During the analysis of such a first law, we, however, come across certain features related to the total entropy change of the horizon, which will be discussed thoroughly.

\section{General Structure of Physical Process First Law and the law for Arbitrary Horizon Cross-sections}
\label{ppfl}
 
Recall that in our earlier analysis, we had imposed a condition while deriving PPFL, that no caustic is formed to the future of the bifurcation surface. This assumption is necessary because one has to integrate the Raychaudhuri equation from the bifurcation surface to some stationary surface in the asymptotic future. However, one can always obtain a first law between two arbitrary horizon slices at the cost of an additional term in the first law.  We begin with a brief outline of the modified PPFL as obtained in \cite{Chakraborty:2017kob}. Consider the horizon perturbation due to some matter influx across the horizon. The entropy associated with the horizon is given by,

\begin{equation}
S = \frac{1}{4} \int _{\mathcal{H}} \sqrt{h} \ d^nx,
\end{equation}
 where the integration is over the horizon cross-section, and $h$ denotes the determinant of the induced metric on its cross-section. The change in entropy along the horizon is given by,
 \begin{equation} \label{delta_S}
 \Delta S = \frac{1}{4} \int d^nx \int_{\lambda_1}^{\lambda_2} d\lambda \ \sqrt{h} \ \theta^{\text{aff}},
 \end{equation}
where the expansion ($\theta^{\text{aff}}$) along the affine parameter $\lambda$ is $\theta^{\text{aff}} = \frac{d}{d \lambda} \ln \sqrt{h}$. One can integrate Eq. (\ref{delta_S}) by parts to obtain,
\begin{eqnarray}
 \Delta S  = \Delta \Bigg[ \frac{1}{4} \int d^nx \ \lambda \ \theta^{\text{aff}}  \sqrt{h}\Bigg]- \frac{1}{4}\int d^nx \int_{\lambda_1}^{\lambda_2}\ d \lambda \ \lambda \ \sqrt{h} \left(\frac{d \theta^{\text{aff}} }{d \lambda}\right).
\end{eqnarray}
This relation can be recast in non-affine parametrization of the geodesics as,
\begin{small}
\begin{eqnarray}\label{entropy_change}
 \Delta S &=&  \Delta \Bigg[ \frac{1}{4} \int d^nx \ \frac{1}{\kappa} \ \theta \sqrt{h}\Bigg] - \frac{1}{4}\int d^nx \ \int_{t_1}^{t_2} \ dt \frac{1}{\kappa} \ \sqrt{h} \left(- \kappa \theta \ + \ \frac{d \theta}{dt}\right),
 \end{eqnarray}
  \end{small}
 where $\kappa$ is the same quantity as defined in section (\ref{formalism}). The affine and the non-affine parameters are related by ${d \lambda}/{dt} = e^{\kappa t}$, and the relation between the expansions  in different parametrizations is given by $\theta^{\text{aff}} = {\theta}/{\kappa \lambda}$. The evolution of $\theta$ along $t$ is governed by the Raychaudhuri equation given by Eq. (\ref{Ray}). We assume that the horizon perturbation is weak so that the higher-order terms can be neglected. Let the tidal field, which sources the evolution of shear is of the order of, $\epsilon = m\kappa$. Truncating Eq. (\ref{Ray}) to lowest order in $\epsilon$, we get
\begin{equation}
-\frac{d \theta}{dt} + \kappa \theta = \mathcal{S}(t) + \sigma^2 \ + \ O(\epsilon^2).
\end{equation}
Consequently, the entropy change (Eq. (\ref{entropy_change})) becomes,
\begin{equation} \label{entrpy_change_2}
\Delta S =  \Delta \Bigg[ \frac{1}{4} \int d^nx \ \frac{1}{\kappa} \ \theta \sqrt{h}\Bigg]  + \frac{1}{4}\int d^nx \  \int_{t_1}^{t_2} dt \ \frac{1}{\kappa}\ \left( \sigma^2 \ + \ \mathcal{S}(t) \right) \sqrt{h}.
\end{equation}

We are interested in the dynamics of those geodesics which do not intersect the infalling body. Hence,  the $\sigma^2$ is the leading order term in this case. The above expression is called the \textit{modified version of the physical process first law}. The important point here is that the above relation determines the entropy change of the horizon between two arbitrary horizon Killing times. However, this relation holds only for some duration of the horizon evolution at which one can neglect the effect of $\theta^2$ terms in the Raychaudhuri equation. The above expression, when evaluated for $t_2\rightarrow \infty$ and $t_1\rightarrow -\infty$, the first term vanishes, and one gets back the original form of the PPFL. The following assumptions are made in the process:
\begin{enumerate}

\item The horizon eventually settles down to a new stationary state. Therefore, the first term evaluated at $t_2\rightarrow\infty$ is zero, by assumption. 

\item The first term is of higher-order at the bifurcation surface, provided the radius of the perturbing body satisfies the bound Eq. (\ref{first_condition}). This bound also ensures that no caustic forms to the future of the bifurcation surface.
\end{enumerate}
We will show that the first assumption does not hold for Rindler horizons in four space-time dimensions. In fact, there is a non zero contribution coming from the upper limit. We will show that though $\theta$ goes to zero at the asymptotic slice, its integral over the cross-section does not. Further, the second term in Eq. (\ref{entrpy_change_2}) diverge when the spatial integration is taken over the entire cross-section. Hence one is bound to define a PPFL for arbitrary cross-sections.\\ 

We will also relax the second assumption in the following sense. The horizon evolution should be quasi-stationary (i.e., no caustic forms) between the concerned horizon slices only. Exploiting our results, one can choose the interval of horizon evolution where such a modified PPFL will hold for a given perturbing object. This way of looking at the problem is different since earlier we were concerned about the size of the perturbing object to keep the horizon evolution quasi-stationary. But now, we look for a suitable interval of horizon Killing time where the first law remains valid. As a consequence, $t_1$ cannot be taken at $-\infty$ but must be some finite value. Hence, we ask, given a model of perturbation can one find the temporal span of the horizon evolution where the entropy change satisfies the relation given by Eq. (\ref{entrpy_change_2})? \\

We answer the question posed above as follows: Suppose an object falls freely across the horizon at $t=0$, which generates a non-zero expansion of the geodesics for $t<0 $. This means that the horizon expands from $t = -\infty$ and ceases to expand when the object hits the horizon ($t=0$). A careful analysis of the Raychaudhuri equation reveals that if $\theta \approx 2 \kappa$, then one cannot neglect the effect of $\theta^2$ term while calculating the entropy change. Now, our problem reduces to finding the horizon Killing time ($t=\tau$) at which $ \theta \approx 2 \kappa$. One can explicitly show that Eq. (\ref{entrpy_change_2}) holds for $t$ satisfying the condition: $\tau < t < \infty$. Consider the perturbation of a Rindler horizon by a spherically symmetric object. Suppose the object crosses the horizon at $t=0$. Using the expressions for shear and expansion, we estimate the time at which the expansion becomes $2 \kappa$ as,

%

\begin{equation}\label{Hbound}
\tau = \frac{1}{\kappa}\ln \Bigg[\frac{1}{2}\left(1-\sqrt{1-N^{-1}}\right)\Bigg],
\end{equation}
where,
\begin{equation}
N = 8 n(n-1)\left( \frac{ \pi m }{\Omega_{n-1}\kappa \rho^n}\right)^2.
\end{equation}

 Once we find the lower bound for the Killing time, we use Eq. (\ref{entrpy_change_2}) to calculate the entropy change for $\tau < t <\infty$. However, one must ensure that the integrations are finite.  This will answer whether the first assumption is correct and how one has to use the general expression  (Eq. (\ref{entrpy_change_2})) to get a finite answer for the entropy change. In the next section, we will show that if the integration over the cross-sections is taken over $[a,\infty)$ (where $a$ should be taken as the radius of the object), then the future slice should be taken at some finite Killing time.

\subsection{Divergences in Four space-time dimension}

 In this section, we evaluate the terms in Eq. (\ref{entrpy_change_2}) separately. Note that the approximations made in the previous section essentially hides all the far region behavior of the perturbing Weyl tensor. However, we find that the shear and the expansion can be calculated from the exact profile, i.e., Eq. (\ref{exactweyl}) of the Weyl tensor, without the assumptions made in the above section. Therefore, we can find the large distance behavior of both the shear and expansion, which are ignored while trying to find a lower bound on the size of the object. We discuss all these features one by one and explain its impact on the validity of the physical process first law.  Now, with the full profile, the horizon shear is given by,

\begin{gather}
\sigma_{\rho\rho}(t) = -\frac{1}{\rho^2} \sigma_{\theta \theta} =\frac{ 2 \kappa  m z_0 e^{\kappa  t}}{\rho ^2} \left[\frac{\bigg( {z_0} \left(e^{\kappa  t}-1\right) \left(3 \rho ^2+2 {z_0}^2 \left(e^{\kappa  t}-1\right)^2\right)\bigg)}{2 \left(\rho ^2+{z_0}^2 \left(e^{\kappa  t}-1\right)^2\right)^{3/2}}-1\right].
\end{gather}
The corresponding expression for expansion of the horizon is,
\begin{gather}\nonumber
\theta(t) = \frac{\kappa  m^2 z_0 e^{\kappa  t}}{8 \rho ^4} \Bigg[-15 \pi  \rho +\frac{64 \left(\rho ^2+2 z_0^2 \left(e^{\kappa  t}-1\right)^2\right)}{\sqrt{\rho ^2+z_0^2 \left(e^{\kappa  t}-1\right)^2}}+30 \rho  \tan ^{-1}\left(\frac{z_0 \left(e^{\kappa  t}-1\right)}{\rho }\right) \\  -\frac{2 z_0 \left(e^{\kappa  t}-1\right) \left(47 \rho ^4+64 z_0^4 \left(e^{\kappa  t}-1\right)^4+113 \rho ^2 z_0^2 \left(e^{\kappa  t}-1\right)^2\right)}{\left(\rho ^2+z_0^2 \left(e^{\kappa  t}-1\right)^2\right)^2}
\Bigg].
\end{gather}

Note that these quantities don't vanish immediately after the perturbing object crosses the horizon \footnote{The horizon expansion and shear become zero just after the perturbing object crosses the horizon if we approximate the electric part of the Weyl
tensor by a delta function as used in  \cite{Amsel:2007mh, Thorne, Bhattacharjee:2014eea}. As mentioned above, we do not make such approximation in this section.}. However, both the shear and the expansion vanishes at $t= \infty$. This ensures that the final black hole configuration is stationary. However, at $\rho=0$, these quantities diverge. If the particle were replaced by an extended object, then these divergences would not have appeared because one should consider the metric in the interior of the body as well. Since we are interested in the horizon generators which do not intersect the body, our range of integration would be from some non-zero value of $\rho$. Hence this divergence is not a problem. Alternatively, one can say that the lower bound on the particle size takes care of this divergence.\\

To write Eq. (\ref{entrpy_change_2}), we need to evaluate the integrations over the cross-sections. Our goal is to verify whether the expression for the entropy change is finite or not. We go ahead and calculate the time and $\rho$ integral of these quantities which appear in Eq. (\ref{entrpy_change_2}). One can show that the contribution to the entropy change due to the first term in Eq. (\ref{entrpy_change_2}) is finite. Hence, we will be interested in the quantity,
\begin{equation}
 \int\big(\Sigma(t_2)-\Sigma(t_1)\big)\rho~d\rho ~d \theta,
\end{equation}
where $\Sigma(t)$ is defined as,
\begin{equation}
\Sigma(t) =  \int dt \  \sigma^2 \ \sqrt{h}.
\end{equation}
 To demonstrate the divergences we will first find this integral over a finite range $\rho~\epsilon~[a,b]$. This is given by the following expression.
\begin{small}
\begin{gather}\label{finite_entropy_chge}
 \int_a^b\big(\Sigma(t_2)-\Sigma(t_1)\big)\rho~d\rho ~d \theta=\frac{\pi \kappa  m^2}{2}  \left[ 16 \log\frac{\left(\sqrt{a^2+z_0^2z_2^2}+z_0 z_2\right)\left(\sqrt{b^2+z_0^2z_1^2}+z_0 z_1\right)}{\left(\sqrt{a^2+z_0^2z_1^2}+z_0 z_1\right)\left(\sqrt{b^2+z_0^2z_2^2}+z_0 z_2\right)} \right.\\ \nonumber
\left.+ z_0^2 z_2(z_2+1)\left(\frac{1}{b^2+z_0^2 z_2^2}- \frac{1}{a^2+z_0^2 z_2^2}\right) - z_0^2 z_1(z_1+1)\left(\frac{1}{b^2+z_0^2 z_1^2}- \frac{1}{a^2+z_0^2 z_1^2}\right) \right.\\ \nonumber 
\left.+ 7\log\frac{\left(a^2+{z_0}^2 {z_1}^2\right)\left(b^2+{z_0}^2 {z_2}^2\right)}{\left(a^2+{z_0}^2 {z_2}^2\right)\left(b^2+{z_0}^2 {z_1}^2\right)}+\bigg(16 {z_0}^2 ({z_1}-{z_2}) ({z_1}+{z_2}+2)\bigg)\left(\frac{1}{b^2}-\frac{1}{a^2}\right) \right.\\ \nonumber
\left. + 16 z_0 \Bigg\{ (z_2+2)\left(  \frac{\sqrt{b^2+z_0^2z_2^2}}{b^2}-\frac{\sqrt{a^2+z_0^2z_1^2}}{a^2}  \right) + (z_1+2)\left( \frac{\sqrt{a^2+z_0^2z_2^2}}{a^2} - \frac{\sqrt{b^2+z_0^2z_1^2}}{b^2} \right)\Bigg\}  \right. \\\nonumber
\left. +\frac{15 z_0}{b}\bigg\{ \tan ^{-1}\left(\frac{{z_0} {z_2}}{b}\right)- \tan ^{-1}\left(\frac{{z_0} {z_1}}{b}\right)\bigg\} +\frac{15 z_0}{a}\bigg\{ \tan ^{-1}\left(\frac{{z_0} {z_1}}{a}\right)- \tan ^{-1}\left(\frac{{z_0} {z_2}}{a}\right)\bigg\}\right],
\end{gather}
\end{small}
where $z_{1,2}=-1+e^{\kappa t_{1,2}}$.
If one takes the future slice at $t_2\rightarrow \infty$ and then takes $b\rightarrow\infty$ limit, then the integral diverges, which means that if one wants to write down a physical process first law for an asymptotic slice besides integrating over the complete slices, then it fails. This was pointed out in \cite{Jacobson:2003wv}. There is, of course, a way to avoid this problem. It is possible to get a finite result ($\Delta S$) with an asymptotic slice if one restricts the change of area to some open region of the horizon slice. We will demonstrate this and explain the consequences. If one takes $t_2\rightarrow \infty$ and subsequently does the $\rho$ integration for some  finite $a, b$, then the above expression (Eq. (\ref{finite_entropy_chge})) becomes,
\begin{gather}\nonumber
\int_a^b\big(\Sigma(\infty)-\Sigma(t_1)\big)\rho~d\rho ~d \theta=\frac{\pi \kappa  m^2}{4}  \left[ -\frac{2 {z_0^2} {z_1} ({z_1}+1) (a-b) (a+b)}{\left(a^2+{z_0}^2 {z_1}^2\right) \left(b^2+{z_0}^2 {z_1}^2\right)} +15 \pi z_0\left(\frac{1 }{b}-\frac{1 }{a}\right)+14 \log \frac{\left(a^2+{z_0}^2 {z_1}^2\right)}{\left(b^2+{z_0}^2 {z_1}^2\right)}\right.\\\nonumber
\left. +32 z_0({z_1}+2)\left(\frac{ \left({z_0} {z_1}-\sqrt{b^2+{z_0}^2 {z_1}^2}\right)}{b^2}-\frac{ \left({z_0} {z_1}-\sqrt{a^2+{z_0}^2 {z_1}^2}\right)}{a^2}\right) \right.\\
\left.  +32 \log\frac{\left(\sqrt{b^2+{z_0}^2 {z_1}^2}+{z_0} {z_1}\right)}{\left(\sqrt{a^2+{z_0}^2 {z_1}^2}+{z_0} {z_1}\right)} + 30 {z_0} \left\{\frac{\tan ^{-1}\left(\frac{{z_0} {z_1}}{a}\right)}{a}-\frac{\tan ^{-1}\left(\frac{{z_0} {z_1}}{b}\right)}{b}\right\} \right].
\end{gather}
 One can see that this quantity is finite for some finite value of $b$, but has a logarithmic term which diverges as $b\rightarrow\infty$. One would think that this can be avoided by putting an upper bound on the size of the particle. This, however, remains to be seen and will be addressed in some future work. Another way to obtain a finite value for $\Delta S$ is the following. If one keeps the final slice to be at some finite Killing time but integrates over the entire horizon cross-section ($b\rightarrow\infty$), the corresponding expression (Eq. (\ref{finite_entropy_chge})) is again finite.

\begin{gather}\nonumber
\int_a^\infty\big(\Sigma(t_2)-\Sigma(t_1)\big)\rho~d\rho ~d \theta=-\frac{\pi \kappa  m^2 }{2} \left[\frac{16 {z_0}^2}{a^2} \bigg(z_1({z_1}+2)-{z_2} ({z_2}+2)\bigg)-\frac{16 {z_0} ({z_1}+2) \sqrt{a^2+{z_0}^2 {z_1}^2}}{a^2}\right.\\\nonumber
\left.+7 \log \frac{\left(a^2+{z_0}^2 {z_2}^2\right)}{\left(a^2+{z_0}^2 {z_1}^2\right)}+16 \log\frac{\left(\sqrt{a^2+{z_0}^2 {z_1}^2}+{z_0} {z_1}\right)}{ \left(\sqrt{a^2+{z_0}^2 {z_2}^2}+{z_0} {z_2}\right)} -\frac{{z_0}^2 {z_1} ({z_1}+1)}{a^2+{z_0}^2 {z_1}^2}+\frac{{z_0}^2 {z_2} ({z_2}+1)}{a^2+{z_0}^2 {z_2}^2}\right.\\
\left.+\frac{16 {z_0} ({z_2}+2) \sqrt{a^2+{z_0}^2 {z_2}^2}}{a^2}+\frac{15 {z_0}}{a}\bigg( \tan ^{-1}\left(\frac{{z_0} {z_2}}{a}\right) -\tan ^{-1}\left(\frac{{z_0} {z_1}}{a}\right)\bigg)\right].
\end{gather}

We will use this feature to write a modified PPFL. Note that this problem arises only for the case of a Rindler horizon. In the case of a black hole horizon, there is a natural cut-off set by the curvature scale of the background black hole geometry, beyond which the Rindler approximation breaks down. In one of the following sections, we will show that this divergent feature does not arise in higher dimensions, even for Rindler horizons. That explains the necessity to write the first law between arbitrary slices at least in 4 spacetime dimensions. The other motivation for writing such a modified PPFL is, of course, evading the lower bound on the radius of the object. We, therefore, conclude this section with an expression for $\Delta S$ between arbitrary slices but with $\rho$ integration in the range $[a,\infty)$.
\begin{gather}
\Delta S =  \frac{ \pi m^2}{8  a^2} \Bigg[ 7 a^2\log \frac{\left(a^2+z_0^2 z_1^2\right)}{\left(a^2+z_0^2 z_2^2\right)}-16 a^2\log \frac{\left(\sqrt{a^2+z_0^2 z_1^2}+z_0 z_1\right)}{\left(\sqrt{a^2+z_0^2 z_2^2}+z_0 z_2\right)}\\ \nonumber
-16 z_0 z_1 \sqrt{a^2+z_0^2 z_1^2}+16 z_0 z_2 \sqrt{a^2+z_0^2 z_2^2}+16 z_0^2 (z_1-z_2) (z_1+z_2)\\\nonumber
+15 a z_0 \left(2 z_2 \tan ^{-1}\left(\frac{z_0 z_2}{a}\right)-2 z_1 \tan ^{-1}\left(\frac{z_0 z_1}{a}\right)+\pi  (z_1-z_2)\right)\Bigg].
\end{gather}
This expression, however, does not take into account the matter stress-energy tensor of the body itself. This contribution can only be calculated if one considers a realistic body for which the metric and the matter content of the interior is known. Note that if the size of the particle satisfies the bound found in section (\ref{boud_cal}) \footnote{However, note that there is, of course, another bound that the radius must satisfy. The radius cannot be less than the Schwarzschild radius of the perturbing body.}, then the initial slice ($t_1$) can be taken to be at the bifurcation surface, else, $t_1$ must satisfy $t_1 > \tau$, where $\tau$ is given in Eq. (\ref{Hbound}). The upper limit $t_2$ cannot, however, be taken to $\infty$, if the integration over the cross-sections is taken up to $\rho\rightarrow\infty$, for Rindler horizons in four spacetime dimension. This expression, therefore, incorporates relaxation of both the assumptions discussed in section (\ref{ppfl}).

\subsection{Dimensions greater than four}
One can define the first law for the horizon cross-sections in higher dimensions, if one considers the evolution satisfying the assumptions detailed in section (\ref{ppfl}). But, as explained before, while calculating the change in entropy using Eq. (\ref{entrpy_change_2}), one should ensure that the integration does not diverge. Interestingly, in dimensions greater than four, the change in entropy is finite even if one considers the evolution of the generators up to infinite Killing time and over the entire horizon cross-sections (excluding  $\rho=0$). To see this, we analyze the expression for the change in entropy for arbitrary slices in higher dimensions. Similar to the case of 4 dimensions, the first term in the expression of entropy change (Eq. (\ref{entrpy_change_2})) is finite even for large values of both $\rho$ and $t$. But, one should check the second term carefully to look for any divergences. We will explicitly calculate this particular term and show that unlike the case of four dimensions, the change in entropy is finite for dimensions higher than 4. First, we consider the evolution of the Rindler horizon in 5 dimensions. The radial component of the electric part of the Weyl tensor which generates shear is given by, 
\begin{gather}
\epsilon_{\rho\rho}=-\frac{32 \  m \rho ^2 \kappa ^2 {z_0}^2 e^{2 \kappa t}}{3 \pi\bigg(\rho ^2+{z_0}^2 \big(e^{\kappa t}-1\big)^2\bigg)^3}.
\end{gather}
The expression for the horizon shear is obtained as,
\begin{gather}\nonumber
\sigma_{\rho\rho}(t)=\frac{4 \kappa  m z_0 e^{\kappa  t}}{3 \pi  \rho ^3} \left[\frac{\rho z_0 \left(e^{\kappa  t}-1\right) \left(5 \rho ^2+3 z_0^2 \left(e^{\kappa  t}-1\right)^2\right)}{\left(\rho ^2+z_0^2 \left(e^{\kappa  t}-1\right)^2\right)^2}+ 3 \tan ^{-1}\left(\frac{z_0 \left(e^{\kappa  t}-1\right)}{\rho }\right)\right]\\  -\frac{2 \kappa z_0 m e^{\kappa  t} }{\rho ^3}.
\end{gather}
Now, one can calculate the expansion of geodesics using Eq. (\ref{theta}).
\begin{gather}\nonumber
\theta(t) = \frac{2 \kappa  m^2 z_0 e^{\kappa  t}}{9 \pi ^2 \rho ^6} \Bigg[-\frac{6 \rho ^2 \left(12 \pi  \rho +z_0 \left(e^{\kappa  t}-1\right)\right)}{\rho ^2+z_0^2 \left(e^{\kappa  t}-1\right)^2}+\frac{16 \rho ^6 z_0 \left(e^{\kappa  t}-1\right)}{\left(\rho ^2+z_0^2 \left(e^{\kappa  t}-1\right)^2\right)^3}-54 \pi ^2 z_0 e^{\kappa  t}\\ \nonumber
+6 \left\{-37 \rho +\frac{24 \rho ^3}{\rho ^2+z_0^2 \left(e^{\kappa  t}-1\right)^2}+36 \pi  z_0 \left(e^{\kappa  t}-1\right)\right\} \tan ^{-1}\left(\frac{z_0 \left(e^{\kappa  t}-1\right)}{\rho }\right)\\ -216 z_0 \left(e^{\kappa  t}-1\right) \tan ^{-1}\left(\frac{z_0 \left(e^{\kappa  t}-1\right)}{\rho }\right)^2
+\frac{68 \rho ^4 z_0 \left(e^{\kappa  t}-1\right)}{\left(\rho ^2+z_0^2 \left(e^{\kappa  t}-1\right)^2\right)^2}+\pi  (37 \rho +18 \pi  z_0)\Bigg].
\end{gather}
Note that the shear and the expansion are zero when evaluated at $t = \infty$, similar to the case of four dimensions. Also, the contribution to the entropy change from the first term of Eq. (\ref{entrpy_change_2}) is finite. Therefore, as in the case of four dimensions, we are worried about the second term in Eq. (\ref{entrpy_change_2}). We explicitly perform the integration to obtain the entropy change. The full expression for  $\int\big(\Sigma(t_2)-\Sigma(t_1)\big)\rho^2~ d\rho ~d\Omega_2$ integrated over $t~\epsilon~[t_1,t_2]$ and $\rho~\epsilon~[a,b]$ can also be found. One would then be able to see that the $t_2\rightarrow\infty$ and the subsequent $b\rightarrow\infty$ gives a finite result,
\begin{gather}\nonumber
\int_a^\infty\big(\Sigma(\infty)-\Sigma(t_1)\big)\rho^2~ d\rho ~d\Omega_2=\frac{4}{9} \pi  \kappa  m^2 \Bigg(-\frac{18 z_0^2 z_1^2}{a^3}+\frac{36 z_0 z_1 (2 a-\pi  z_0)}{\pi  a^3}+\frac{18 \left(\pi ^2-4\right) a+39 \pi  z_0}{\pi ^2 a^2}-\frac{8 \left(a^3-a z_0^2 z_1\right)}{\pi ^2 \left(a^2+z_0^2 z_1^2\right)^2}\\ \nonumber+\frac{72 z_0^2 z_1-60 a^2}{\pi ^2 \left(a^3+a z_0^2 z_1^2\right)}-\frac{6 (2 z_1+1) \cot ^{-1}\left(\frac{z_0 z_1}{a}\right)}{\pi ^2 z_0 z_1^2}-\frac{6}{\pi ^2 a z_1}+\frac{6 z_1+3}{\pi  z_0 z_1^2}\\ \nonumber- \left(12 \left(z_0^2 z_1 (z_1+2)-a^2\right) \tan ^{-1}\left(\frac{z_0 z_1}{a}\right)+a z_0 (24 z_1+13)+12 \pi\left(a^2-  z_0^2 z_1 (z_1+2)\right)\right) \\ \times \left(\frac{z_0 z_1}{a}\right) \frac{6}{\pi ^2 a^3} \tan ^{-1}\Bigg).
\end{gather}
Therefore, in the case of Rindler horizons in five spacetime dimension, the first assumption made in section (\ref{ppfl}) continues to hold. Hence there is no necessity to take the future slice to be at a finite Killing time. However, in principle, one can write an expression for entropy when the future slice is taken at some finite Killing time. The relaxation of the second condition can still be done (section (\ref{ppfl})).  The corresponding expression for entropy change is found to be,
\begin{gather}
\Delta S = \frac{ m^2 }{9 \pi  a^3 z_0 z_1 \left(a^2+z_0^2 z_1^2\right)}\Bigg[z_0 z_1 \Big(2 \left(9 \pi ^2-67\right) a^4+33 \pi  a^3 z_0 z_1 \\ \nonumber +6 \left(6 \pi ^2-23\right) a^2 z_0^2 z_1^2+33 \pi  a z_0^3 z_1^3+18 \pi ^2 z_0^4 z_1^4\Big)+ 6 \left(a^2+z_0^2 z_1^2\right) \tan ^{-1}\left(\frac{z_0 z_1}{a}\right)\\ \nonumber \times \left(a^3+12 z_0 z_1 \left(a^2+z_0^2 z_1^2\right) \tan ^{-1}\left(\frac{z_0 z_1}{a}\right)-12 \pi  a^2 z_0 z_1-11 a z_0^2 z_1^2-12 \pi  z_0^3 z_1^3\right)\Bigg].
\end{gather}

This expression again does not take into account the matter stress-energy tensor of the body itself. The future slice has been taken at the asymptotic Killing time. If $a$ does not satisfy the bound given by Eq. (\ref{Hrbound}), on the radius then $t_1$ must satisfy the bound $t_1>\tau$, where $\tau$ is given by Eq. (\ref{Hbound}). If $a$ satisfies the bound then $t_1$ can be taken at $-\infty$. 

For completeness, we will briefly outline the results obtained for six-dimensional Rindler space-time. In 6 dimensions, we have the following expressions for the electric part of Weyl tensor.
\begin{gather}
\epsilon_{\rho\rho}=-\frac{45 \  m \rho ^2 \kappa ^2 {z_0}^2 e^{2 \kappa t}}{4 \pi\bigg(\rho ^2+{z_0}^2 \big(e^{\kappa t}-1\big)^2\bigg)^{\frac{7}{2}}}.
\end{gather}
The expression for shear is given by,
\begin{gather}
\sigma_{\rho\rho} (t)=\frac{3 \kappa  m z_0^2 e^{\kappa  t} \left(e^{\kappa  t}-1\right) \left(15 \rho ^4+8 z_0^4 \left(e^{\kappa  t}-1\right)^4+20 \rho ^2 z_0^2 \left(e^{\kappa  t}-1\right)^2\right)}{4 \pi  \rho ^4 \left(\rho ^2+z_0^2 \left(e^{\kappa  t}-1\right)^2\right)^{5/2}}-\frac{6 \kappa  m z_0 e^{\kappa  t}}{\pi  \rho ^4}.
\end{gather}
This expression is zero at $t = \infty$. One procceds to calculate the expansion and is found to be zero at $t = \infty$, which ensures the future equilibrium configuration of the horizon. The contribution to the entropy change from the first term of Eq. (\ref{entrpy_change_2}) is finite. Therefore, we calculate the  term  $\int\big(\Sigma(t_2)-\Sigma(t_1)\big) \ \rho^3\ d\rho \ d\Omega_3$ integrated over $t~\epsilon~[t_1,t_2]$ and $\rho~\epsilon~[a,b]$. The $t_2\rightarrow\infty$ and the subsequent $b\rightarrow\infty$ again gives a finite result. The corresponding expression for change in entropy of Rindler horizon in 6 dimensions is calculated to be,
\begin{gather}
\Delta S = \frac{9 m^2}{2048 }  \Bigg[\frac{1890 z_0 z_1 }{a^3}\cot ^{-1}\left(\frac{z_0 z_1}{a}\right)+\frac{2 \left(73 a^2+67 z_0^2 z_1^2\right)}{\left(a^2+z_0^2 z_1^2\right)^2}+\frac{158}{a^2}\\ \nonumber+\frac{4096 z_0 z_1 \left(z_0 z_1-\sqrt{a^2+z_0^2 z_1^2}\right)}{a^4}\Bigg].
\end{gather}
We have analyzed the behavior of the entropy change for a Rindler horizon in different dimensions. An interesting observation thus made is, in 4 dimensions the expression for entropy change is infinite if one considers the evolution of all the geodesics for $\rho~\epsilon~(0,\infty]$, and  $t \rightarrow \infty$ limit. But such divergences do not arise for dimensions five and six. Therefore, by looking at the behavior of the expression for the entropy change, we conjecture that the change in horizon entropy is finite even the horizon evolution is taken up to $t \rightarrow \infty$ for dimensions greater than 4. We will analyze the same problem for space-times with non-zero cosmological constant below.\\

\section{Horizon perturbation in Arbitrary dimensional de-Sitter space-time}\label{desitter_}
As the final step of our analysis on the validity of PPFL, we consider the perturbation of a black hole horizon in $ (n+2)$ dimensional spacetime with a non-zero cosmological constant ($\Lambda$). Such space-times demand similar studies in their own rights owing to their importance in cosmology and holography. The interesting question now is how the introduction of a cosmological constant changes our previous results. We will be dealing with de-Sitter space-time only since the examination with the anti-de-Sitter is quite similar. In a small enough region, the event horizon of a Schwarzschild-de Sitter black hole 
can be approximated by a horizon, as perceived by an accelerating observer in Minkowski spacetime (Appendix \ref{one}). We shall exploit this advantage to simplify our calculations. We will be considering only the perturbation of the black hole horizon ($r_h$) throughout this work since the cosmological horizon is of less importance for the study we are having. Basically, the calculations deal with the perturbation of the Rindler horizon in flat-space-time and the effect of $\Lambda$ comes from the horizon perturbing matter. At the end of the calculations, we will revert the results back to the case of Schwarzschild-de Sitter black hole, as explained in Appendix (\ref{one}). We follow the same procedure as before to find the expansion of the horizon due to an infalling spherically symmetric object. We assume the mass $m$ of the perturbing object is small so that one can treat the problem perturbatively. This allows us to consider the perturbing metric as a solution of linearized Einstein's equations in the presence of a positive cosmological constant.  In isotropic coordinate system $(T,Z, \rho,\Omega_{n-1})$, the metric for the perturbing object is given by \cite{Astefanesei:2003gw},
\begin{small}
\begin{gather}
ds^2 = -\left(1-\frac{C m e^{-(n-1)H_0T}}{ \left(\sqrt{\left(\rho^2+(Z-{z_0})^2\right)}\right)^{(n-1)}}\right) dT^2 \\\nonumber
+ e^{2 H_0 T}\left( 1+ \frac{C m e^{-(n-1)H_0T}}{ (n-1)\left(\sqrt{\rho^2+(Z-z_0)^2}\right)^{(n-1)}}\right)\left(dZ^2+ d\rho^2 + \rho^2 d\Omega_{n-1}^2\right),
\end{gather}
\end{small}
where $ \rho^2 = \sum_{i=1}^n x_i^2$ and $H_0^2 = 2 \Lambda /(n(n+1))$. The constant $C = \frac{16 \pi}{n \Omega_n}$, with $ \Omega_n$ being the volume of $S^n$. The non-zero components of the electric part of the Weyl tensor are calculated  along the horizon ($Z-T=0$).
\begin{small}
\begin{gather}\label{Weyl_1}
\varepsilon_{\rho \rho} =  - \frac{(n-1)}{g_{\theta_i \theta_i}}\varepsilon_{\theta_i \theta_i}\\ \nonumber = -\frac{ (n-1)C m \kappa^2 Z^2 e^{-(n-1)H_0 Z}}{2\left(\sqrt{\rho^2+(Z-z_0)^2}\right)^{(n+3)}} 
\Bigg(\rho^2\left(n+e^{2 H_0 Z}\right) + (Z-z_0)^2\left(e^{2 H_0 Z}-1\right)\Bigg).
\end{gather}
\end{small}
Proceeding with the same analysis as before, we approximate the electric part of the Weyl tensor with a delta function centered at $\bar{t} = 0$, where $\bar{t}$ is the shifted Killing time.
\begin{gather}
\varepsilon_{\rho \rho} (\bar{t})=  - \frac{(n-1)}{g_{\theta_i \theta_i}}\varepsilon_{\theta_i \theta_i}= -\frac{(n-1)}{n\Omega_{n-1}} \frac{8 \pi m \kappa z_0 }{\rho^n} e^{-(n-1)H_0 z_0}\left(n-1+e^{2H_0 z_0}\right) \delta(\bar{t}).
\end{gather}
This expression reduces to the one obtained in \cite{Amsel:2007mh} for $H_0 =0$. We calculate the shear of the horizon using Eq. (\ref{sigma}) and the expansion using Eq. (\ref{theta}).

\begin{gather}
\theta(\bar{t}) = \frac{n}{(n-1)\kappa}\left(\frac{(n-1)}{n\Omega_{n-1}} \frac{8 \pi m \kappa z_0 }{\rho^n} e^{-(n-1)H_0 z_0}\left(n-1+e^{2H_0 z_0}\right)\right)^2 \\ \nonumber \times e^{\kappa \bar{t}}\left(1-e^{\kappa \bar{t}}\right)\Theta(-\bar{t}).
\end{gather}
 The condition on $\theta$ such that a caustic forms, now reads,
 \begin{equation}
\left(\frac{\theta}{n \kappa}\right)_{max} \geq 1.
\end{equation}
This gives a limit on the size of the perturbing object to keep the horizon evolution quasi-stationary. Therefore, caustic formation along the horizon is avoided if,
\begin{gather}\label{final_1}
r > 
 \left(\frac{4 \pi\sqrt{n-1} m z_0 }{n \Omega_{n-1}}  e^{-(n-1)H_0 z_0}\left(n-1+e^{2H_0 z_0}\right)\right)^{\frac{1}{n}},
\end{gather}
where we have approximated the radial coordinate $\rho$ with the radius of the object. One can check that for  $\Lambda = 0$; this condition reduces the expression obtained in Eq. (\ref{Hrbound}). For the case of black holes, the condition on the radius of perturbing matter to avoid caustic formation can be obtained as described in Appendix (\ref{one}). One can see from the above expression that the allowed (no caustic) values for the radius of the perturbing object get larger as the dimensionality increases.\\
\\
Our analysis on the validity of PPFL for various perturbation set-ups completes here. We have investigated different cases of horizon perturbations and obtained conditions on the size of the object so that caustic will not form along the horizon, hence ensuring the validity of the first law. In the next section, we consider the case of modified PPFL relation discussed in \cite{Chakraborty:2017kob}, where authors have obtained a first law relation for arbitrary horizon cross-sections. In the next section, we analyze the characteristics of the entropy change for a space-time with non-zero cosmological constant.

\subsection{PPFL for arbitrary horizon cross sections in de-Sitter space-time}\label{EC}
In this section, we look for a duration of horizon evolution in de-Sitter space-time, where one can establish the first law even if caustics form. As discussed before, caustic may develop along the horizon if the radius of the perturbing object doesn't obey Eq. (\ref{final_1}). In the case of the horizon perturbation in Schwarzschild space-time, we have already explained the appropriate way to develop the notion of the first law in section (\ref{ppfl}). Now, we develop the same in the case of black hole perturbation in de-Sitter space-time, exploiting the results obtained above.

Before proceeding, let us recap the motive behind considering the horizon evolution between arbitrary slices. We have explained in section (\ref{formalism}) that if the expansion of the black hole horizon becomes comparable to $2\kappa$, where $\kappa$ is the surface gravity of the black hole before perturbation, due to infall of some matter at time $t=0$, caustic will inevitably form along the horizon at some finite earlier time ($t<0$). Evidently, the quasi-stationary approximation of the horizon evolution breaks down hence invalidating the first law. Interestingly, it has been shown in \cite{Chakraborty:2017kob}, that one can still study the horizon evolution by excluding the non-quasi-stationary part of the evolution. This is achieved by slightly modifying the form of the first law as in section (\ref{ppfl}). Also, we have explained how this works in the case of horizon perturbation of the Schwarzschild black hole. Here we give yet another simple calculation to exemplify the modified version of PPFL.

Consider the perturbation of a $(n+2)$ dimensional Rindler horizon in de-Sitter spacetime by a spherically symmetric object. Suppose the radius of the object violates the condition to avoid caustic formation ($\theta < n \kappa$), at some point of horizon evolution ($t=\tau$). This spoils the quasi-stationarity nature of the process for $t < \tau$. But, one can still study the first law for $\tau < t < \infty$. The value of $\tau$, for this case, can be calculated as,
\begin{equation}
\tau = \frac{1}{\kappa}\ln \Bigg[\frac{1}{2}\left(1-\sqrt{1-N^{-1}}\right)\Bigg],
\end{equation}
where,
\begin{equation}
N =  \frac{8 (n-1)}{n}\left( \frac{ \pi m z_0}{\Omega_{n-1} \rho^n} e^{-\frac{(n-1)H_0}{\kappa}}\left(n-1+e^{\frac{2 H_0}{\kappa}}\right)\right)^2.
\end{equation}

 One can safely use Eq. (\ref{entrpy_change_2}) to obtain the change in entropy within the range $\tau < t <\infty$. Also, it will be interesting to check whether the change in entropy is finite. Since the calculation of area change for arbitrary dimensional de Sitter space-time is quite laborious, we consider four (4D) as well as six-dimensional (6D) cases only. The expansion and shear calculated for 4D and 6D are well behaved except when $\rho$ becomes zero, just like the case of Rindler horizon in flat space-time. This divergence is, however, not our concern since we are looking for the evolution of those geodesics which do not intersect the perturbing object. Further, the first term in the expression of entropy change Eq. (\ref{entrpy_change_2}) is finite for both 4D and 6D cases even if one considers the $\rho \rightarrow \infty$ limit. Hence, just like the case of flat space-time, divergences come from the second term of Eq. (\ref{entrpy_change_2}). In 4D, we denote this term as $\int\big(\Sigma(t_2)-\Sigma(t_1)\big)\rho d\rho d \theta$, where $\Sigma(t)$ is the time integral of $\sigma^2$. This quantity is finite if either one of the variables ($\rho$ or $t_2$) varies till infinity, but other is held finite. But diverges if both tend to infinity. This can be seen by the following argument. If one evaluates the above integration over the cross-section area, where $\rho$ varies between $[a,b]$, then an asymptotic expansion around $t_2 = \infty$ gives the following expression:
 \begin{gather} \nonumber
\int_a^b\big(\Sigma(\infty)-\Sigma(t_1)\big)\rho d\rho d \theta = \frac{\pi  \kappa  m^2 e^{-2 H_0 z_0}}{4 a^2 b^2} \Bigg[z_0 (a-b) \Bigg\{a \left(\pi  b \left(8 e^{2 H_0 z_0}+8 e^{4 H_0 z_0}-1\right)-8 z_0 \left(e^{2 H_0 z_0}+1\right)^2\right)\Bigg\} \\ \nonumber
-32 a^2 b^2 e^{2 H_0 z_0} \log \left(\frac{a}{b}\right) \cosh (H_0 z_0) \Bigg\{\cosh (H_0 z_0)+2 \Bigg(\log (2z_0)-1\Bigg) \sinh (H_0 z_0)\bigg\} \\  -8 b z_0^2 (a-b) \left(e^{2 H_0 z_0}+1\right)^2\Bigg],
 \end{gather}
 as the leading order term. However, this expression contains $\log{b}$ term and therefore diverges when the $b\rightarrow\infty$ limit is taken. But if we consider higher-dimensional de-Sitter spacetimes ($D \geq 5 $), this divergence goes away, giving finite results for entropy change, similar to the case of Rindler horizon in flat space-time. The expression calculated for the change in entropy of Rindler horizon in six dimensions, obtained by taking both $t\rightarrow \infty$ and $\rho\rightarrow \infty$ limits is given,
 
 \begin{gather}\nonumber
\Delta S= \frac{9   m^2 e^{-4 H_0 z_0} }{2048 } \Bigg[\frac{5 z_0 z_1 \left(16 e^{2 H_0 z_0} \left(e^{2 H_0 z_0}+5\right)+93\right) \left(\pi -2 \tan ^{-1}\left(\frac{z_0 z_1}{a}\right)\right)}{a^3}\\ \nonumber+\frac{1}{z_0^2 z_1^2}\Bigg\{\frac{2 z_0 z_1}{a^4 \left(a^2+z_0^2 z_1^2\right)^{5/2}} \Bigg(128 z_0^7 z_1^7 \left(e^{2 H_0 z_0}+3\right)^2 \left(\sqrt{a^2+z_0^2 z_1^2}-z_0 z_1\right)\\ \nonumber -64 a^8 \left(2 e^{2 H_0 z_0}+e^{4 H_0 z_0}-3\right)+a^2 z_0^5 z_1^5 \Big(\left(1648 e^{2 H_0 z_0}+304 e^{4 H_0 z_0}+2223\right) \sqrt{a^2+z_0^2 z_1^2}\\ \nonumber -64 z_0 z_1 \left(e^{2 H_0 z_0}+3\right) \left(7 e^{2 H_0 z_0}+17\right)\Big)+a^4 z_0^3 z_1^3 \Big(-192 z_0 z_1 \left(e^{2 H_0 z_0}+3\right) \left(3 e^{2 H_0 z_0}+5\right)\\ \nonumber  +\left(112 e^{2 H_0 z_0} \left(2 e^{2 H_0 z_0}+9\right)+1041\right) \sqrt{a^2+z_0^2 z_1^2}\Big)-8 a^6 z_0 z_1 \Big(8 z_0 z_1 \left(e^{2 H_0 z_0}+3\right) \\ \nonumber \times \left(5 e^{2 H_0 z_0}+3\right) + \left(-16 e^{2 H_0 z_0}-6 e^{4 H_0 z_0}+3\right) \sqrt{a^2+z_0^2 z_1^2}\Big)\Bigg)\\ -64 \left(2 e^{2 H_0 z_0}+e^{4 H_0 z_0}-3\right) \left(\log \left(a^2+z_0^2 z_1^2\right)-2 \log \left(\sqrt{a^2+z_0^2 z_1^2}+z_0 z_1\right)\right)\Bigg\}\Bigg].
 \end{gather}
 One can see that the above expression is finite. Therefore, for a Rindler horizon in a space-time with non-zero cosmological constant, the change in horizon entropy, when considered between asymptotic cross-sections, is finite for dimensions greater than four. This result is the same as of a Rindler horizon in flat space-time. \\

Finally, we would like to emphasize the usefulness of the recipe for obtaining a period of horizon evolution in which the process is quasi-stationary. As presented in \cite{Chakraborty:2017kob}, one doesn't always have to worry about the bifurcation surface to establish the first law for the black hole horizon. Instead, it can be achieved by considering any arbitrary horizon slices, provided the quasi-stationarity approximation is valid in between such cross-sections, and the first law looks like Eq. (\ref{entrpy_change_2}). We have explicitly calculated the value of the time slice in terms of the parameters of the perturbing object.
  
\section{discussion}
The first law of black hole mechanics has moulded much of our views about the black holes. However, checking the limits of application of the first law of black hole mechanics is an important aspect that can be explored. As emphasized before, the questions posed in this paper are analogous to the question of which processes are quasi-static in usual thermodynamics. However, there is an important difference. In usual thermodynamics, if one starts with an initial equilibrium state, then a non-quasi-static process will evolve the state through non-equilibrium states. Due to the teleological nature of the event horizon one has to ask the opposite question in the case of black hole thermodynamics. That is if one assumes that the final state is a stationary black hole, what are the processes that ensure that the initial state was stationary. The break down of the assumptions in PPFL and the onset of caustics to the past, of the event of an object crossing the horizon, precisely means that the initial state was non-stationary and that the initial bifurcation surface never existed.\\

While this has been pointed out several times in the literature, it has also been argued that Rindler horizons present itself in a unique way, due to non-compact cross-sections. In \cite{Jacobson:2003wv}, it has been pointed out that though the shear remains small throughout the process, the total shear obtained by integrating over the cross-sections contribute infinitely to the change in entropy. In this article, the main focus is to demonstrate this precisely. In four dimensions, we show, that this indeed is true. However, in higher dimensions, this is not so, at least for the case when the perturbation is due to a spherical body. We attribute this behavior to the fall-off condition of the Weyl tensor in asymptotically flat space-times in four dimensions. It is possibly similar in spirit to the logarithmic behavior of the electric potential in two dimensions. Assuming that the occurrence of such pathologies is generic; we provide a recipe to obtain finite results by restricting ourself to finding changes only within a finite interval of horizon evolution (modified PPFL). \\


A similar study has been done in the case where the background is de-sitter.  The results obtained have been compared with the case of zero cosmological constant. In effect, we observe that the set of allowed values for the size of the perturbing object, so that the PPFL remains valid, get reduced due to the presence of the cosmological constant. Also, our results show that this effect of the cosmological constant is the same in all dimensions. The motivation for the analysis in arbitrary dimensions is largely due to the growing interest in ``the large dimension'' limits of Einstein's equation and its solutions. The results obtained show that the allowed range of values of the radius of the object monotonically increases as the dimensions increases.\\

Another important aspect is the following. Even if caustics form to the future of the bifurcation surface, it is plausible that the PPFL still holds for an interval time after caustics set in. However, in this case, the limits of integration are different from those assumed in the original derivation of PPFL. The effect of such a choice of integration limits is a modified version of PPFL suggested in   \cite{Chakraborty:2017kob}. This motivates one to find a horizon time beyond which this modified PPFL continues to hold. This is precisely what we have calculated in sections  \ref{ppfl} and \ref{EC}. This lends support and provides an example for the suggestions put forward in \cite{Chakraborty:2017kob}.

\section{Acknowledgements}
We thank Sudipta Sarkar for suggesting the problem and discussions. We also thank Avirup Ghosh for extensive discussions.

\appendix

\section{Rindler approximations of black hole space-times}\label{one}
We outline the idea of the Rindler approximation of black hole horizon, which shows that the near horizon geometry of a stationary non extremal black hole is that a Rindler spacetime. Though we have considered Schwarzschild and Schwarzschild-de Sitter space-times for our case studies, this idea is applicable to more general cases. We consider a general spherically symmetric metric of the form,
\begin{equation}
ds^2 = -f(r) dt^2 +\frac{dr^2}{f(r)} + r^2 d\Omega^2.
\end{equation}
We expand the function $f(r)$ around the black hole horizon as,
 \begin{equation}
 f(r) = f(r_+) + f'(r_+) (r-r_+) + O\left((r-r_+)^2\right),
 \end{equation}
where $r_+$ represents the position of black hole horizon. If we restrict the region of interest sufficiently close to the horizon, one can approximate the above relation as,
\begin{equation}
f(r) \approx 2 \kappa (r-r_+),
\end{equation}
where $ \kappa = \frac{1}{2} f'(r_+)$ is the surface gravity of the corresponding horizon. Introducing a new radial coordinate ($\alpha$) which vanishes at the horizon and increases outwards as,\\
\begin{equation}
\alpha = \sqrt{2 \kappa (r-r_+)} + O\left((r-r_+)^{\frac{3}{2}}\right).
\end{equation}
The line element becomes,
\begin{equation}
ds^2 = -\alpha^2 dt^2 +\frac{d\alpha^2}{\kappa^2} + r_+^2 d\Omega^2.
\end{equation}
 Now one can express the above metric in cartesian coordinates in a small region of space-time sufficiently close to the horizon ($\alpha <<1$) around a specific direction $\theta = \theta_0$ as follows
\begin{gather}
x = r_+ \sin(\theta_0) \phi, \\ \nonumber
y =r_+ (\theta -\theta_0),\\\nonumber
z = \frac{\alpha}{\kappa}.
\end{gather}
The metric in this coordinates will looks like,
\begin{equation}
ds^2 = -\kappa^2 z^2 dt^2 + dx^2+dy^2+dz^2.
\end{equation}
This is the Rindler spacetime and it will be obvious if we do another coordinate transformation as,
\begin{gather}
T = z\sinh(\kappa t) \text{ and } Z = z \cosh (\kappa t).
\end{gather}
And the metric becomes,
\begin{equation}
ds^2 = - dT^2 + dx^2 + dy^2 + dZ^2.
\end{equation}
The Rindler approximation of black hole horizon was justified by the following assumptions. The region under consideration is within $ \alpha <<1$, which means on the trajectory, $ z_0 << \frac{1}{\kappa}$ and is within a  patch of horizon cross-section which is small compared to the total area of the horizon cross-sections. To recover the results for black holes, we relax these approximations so that $ \kappa z_0 = \alpha_0 \approx 1$. Where $\alpha_0$ is the initial radial distance of the perturbing object from the horizon \cite{Suen:1988kq}.

\end{document}